\documentclass{article}

\usepackage{arxiv}

\usepackage[utf8]{inputenc} 
\usepackage[T1]{fontenc}    
\usepackage{hyperref}       
\usepackage{url}            
\usepackage{booktabs}       
\usepackage{amsfonts}       
\usepackage{nicefrac}       
\usepackage{microtype}      
\usepackage{lipsum}
\usepackage{graphicx}
\usepackage{amsmath,amssymb}
\usepackage{bm}

\title{Geometric Transition of Latent Dynamics in Intermittent Chaos: SVD-based Effective Dimension for Detection and Proactive Control}

\author{
  Takashi Hikihara\thanks{Present address: National Institute of Informatics, Tokyo, Japan} \\
  Kyoto University\\
  Yoshida-Honmachi, Sakyo-ku, Kyoto, Japan \\
  \texttt{hikihara.takashi.2n@kyoto-u.ac.jp} \\
  \And
  Tetsushi Saburi \\
  Graduate School of Kyoto University\\
  Katsura, Nishikyo-ku, Kyoto 615-8510, Japan \\
  \texttt{saburi.tetsushi.46z@kyoto-u.jp} \\
}

\begin{document}
\maketitle

\begin{abstract}
This paper investigates the geometric detection of latent dynamic transitions (where the latent space is defined as a finite-dimensional state space reconstructed via a delay-embedding Hankel matrix) in mutually coupled phase-locked loops (PLLs) exhibiting on-off intermittent chaos. We propose a data-driven approach to identify the transition of the effective dimension in latent dynamics as a precursor to physical bursting events. By applying Singular Value Decomposition (SVD) to a Hankel matrix constructed from high-resolution oversampled data, we extract the effective dimension that reflects the dynamical shift of intrinsic variables---from a synchronized two-dimensional manifold to a higher-dimensional chaotic state---prior to the actual burst. This geometric indicator enables a proactive, event-driven intervention using Pyragas-type delayed feedback control, which is activated only when the detected dimension exceeds a critical threshold. Simulation results demonstrate that this geometric precursor effectively compensates for control latencies and substantially reduces accumulated control energy compared to conventional continuous feedback. The findings establish a methodological strategy for stabilizing infinite-dimensional delayed systems by exploiting early warning signals embedded in dimensional transitions of latent dynamics.
\end{abstract}


\section{Introduction}
Synchronization in coupled system architectures is vital for maintaining functional stability across networked infrastructures. In the presence of signal propagation delays, these systems are naturally modeled by delay differential equations (DDEs), whose state spaces are inherently infinite-dimensional. While DDEs possess infinite-dimensional phase spaces, on-off intermittency is fundamentally driven by transverse instability near a chaotic synchronization manifold \cite{Fujisaka1983, Platt1993, Ott1994, Pikovsky2001}. In this phenomenon, two synchronized chaotic trajectories intermittently desynchronize into large-amplitude bursts, rather than being chaotic solely during desynchronization.

Early warning indicators for dynamic transitions in nonlinear systems have conventionally relied on statistical markers associated with critical slowing down (CSD), such as variance expansion or autocorrelation time prolongation \cite{scheffer2009, Leemput2013}. For instability suppression, continuous time-delay feedback architectures---such as Pyragas control \cite{Pyragas1992}---have been applied across various physical and mechanical platforms \cite{Handbook2007, hikihara1996, hikihara1997}. 

Although dynamically switching the mutual coupling strength on and off might appear to be an alternative for maintaining synchronization, modulating coupling parameters continuously alters the system's structural vector field and introduces high-frequency switching shocks into the delay system. In contrast, Pyragas-type delayed feedback acts non-invasively on the transverse phase difference without modifying internal system parameters, naturally vanishing once phase synchronization is restored.

However, conventional continuous feedback requires non-zero control input even during small perturbations, imposing substantial energy costs and altering the autonomous dynamics.

To address these limitations, this study presents a data-driven control strategy that utilizes time-delay embedding and Singular Value Decomposition (SVD) to detect precursor dimensional shifts prior to macroscopic desynchronization \cite{Hiroe2023}. Here, the term ``latent space'' refers specifically to the finite-dimensional state space reconstructed from a scalar observation series via a delay-embedding Hankel matrix. Mapping scalar observations into this latent space yields an effective dimension metric that serves as a predictive trigger for event-driven feedback intervention, eliminating the need for continuous control signals.

\section{System Model and Methodology}

\subsection{Coupled Phase-Locked Loop Model and On-Off Intermittency}
\begin{figure}[htbp]
\centering
\includegraphics[width=0.7\textwidth]{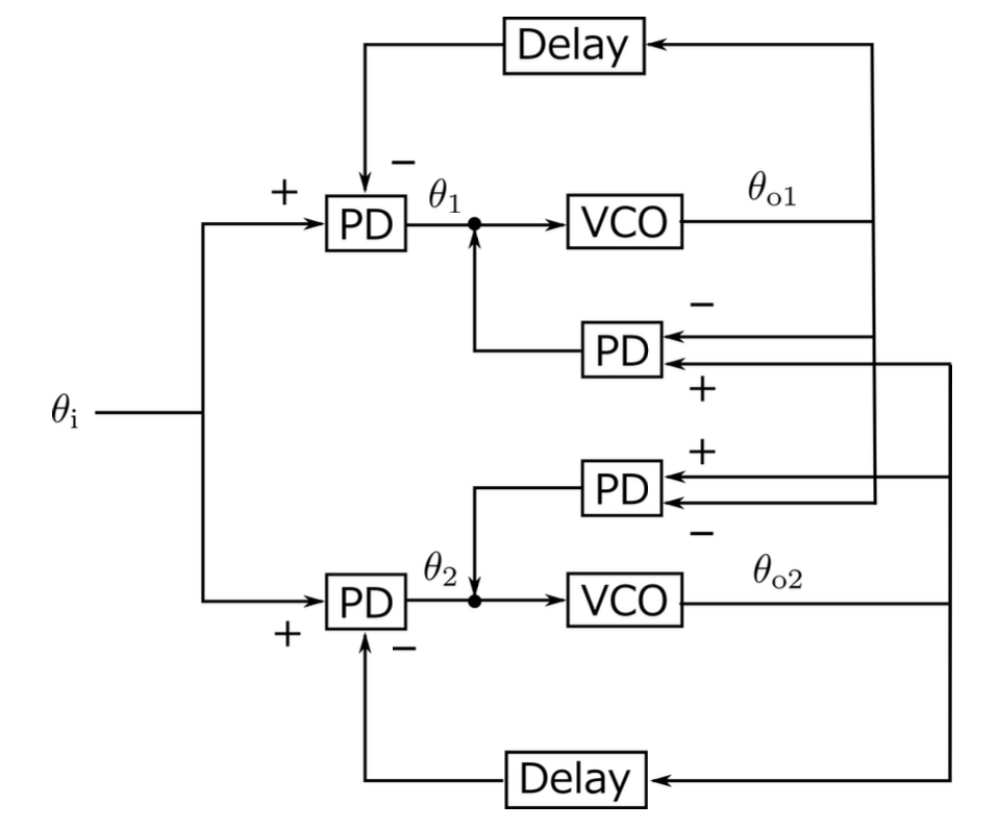}
\caption{Block diagram of mutually coupled phase-locked loops (PLLs) with propagation delay $\tau$. Each loop consists of Phase Detectors (PD), a Voltage-Controlled Oscillator (VCO), and a feedback delay unit. The phase differences $\theta_1(t)$ and $\theta_2(t)$ are derived from the reference phase $\theta_i$ and cross-coupled delayed outputs $\theta_{\rm{o}1}$ and $\theta_{\rm{o}2}$, governing the transverse desynchronization dynamics in Eq.~(1).}
\label{fig:PLLs}
\end{figure}

We consider a system of two mutually coupled phase-locked loops (PLLs) with propagation delay $\tau_d$, as schematically illustrated in Fig.~\ref{fig:PLLs}~\cite{Linsey1976, Weinberg1975, Dessouky1984}. The system is formulated in a rotating frame of reference, where $\delta$ represents the relative natural frequency deviation (detuning) with respect to the fundamental reference frequency $f_0 = 1/T_0$. The fundamental physical timescale is governed by $f_0$, the oscillation period $T_0$, and the transmission delay $\tau_d$. Defining the normalized delay as $\tau = \tau_d / T_0$, the governing delay differential equations (DDEs) for the input-output phase differences $\theta_1(t)$ and $\theta_2(t)$ in the uncontrolled state are given by:
\begin{equation}
\begin{cases}
\dot{\theta}_{1}(t) = \delta - \sin(\theta_{1}(t-\tau)) + g \sin(\theta_{2}(t) - \theta_{1}(t)) \\
\dot{\theta}_{2}(t) = \delta - \sin(\theta_{2}(t-\tau)) + g \sin(\theta_{1}(t) - \theta_{2}(t))
\end{cases}
\end{equation}
where $g$ denotes the mutual coupling strength. To eliminate notation ambiguity throughout this paper, $\tau$ strictly represents the normalized loop delay parameter ($\tau = 5.0$), and $T_{\text{lam}}$ denotes the laminar phase durations in statistical evaluations.

To analyze the transverse stability of the synchronized state ($\theta_1 = \theta_2$), we introduce the average phase $U(t) = (\theta_1 + \theta_2)/2$ and phase difference $V(t) = (\theta_1 - \theta_2)/2$:
\begin{equation}
\begin{cases}
\dot{U}(t) = \delta - \frac{1}{2} [ \sin(U_{\tau} + V_{\tau}) + \sin(U_{\tau} - V_{\tau}) ] \\
\dot{V}(t) = -g \sin(2V(t)) + \frac{1}{2} [ \sin(U_{\tau} - V_{\tau}) - \sin(U_{\tau} + V_{\tau}) ]
\end{cases}
\end{equation}
where $U_{\tau} \equiv U(t-\tau)$ and $V_{\tau} \equiv V(t-\tau)$. The subspace $V(t) = 0$ corresponds to the synchronization manifold. With system parameters set to $g = 0.2125$, $\delta = 0.5$, and $\tau = 5.0$, the transverse manifold loses asymptotic stability, driving the system into the immediate supercritical vicinity of a blowout bifurcation. Note that because both PLLs possess identical natural frequency detuning ($\delta_1 = \delta_2 = \delta = 0.5$) in the rotating reference frame, $\delta$ cancels out in Eq.~(2), ensuring that $V(t) = (\theta_1 - \theta_2)/2$ represents transverse phase deviations without linear phase drift or frequency-mismatch phase jumps.

\begin{figure}[htbp]
\centering
\includegraphics[width=0.5\textwidth]{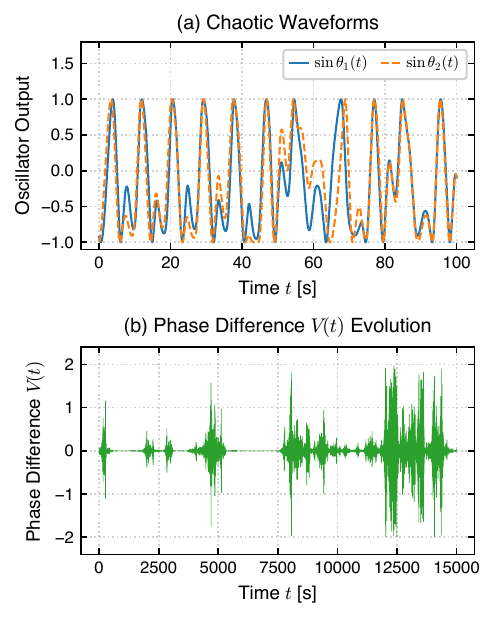}
\caption{Chaotic time-series dynamics exhibiting on-off intermittency in the delay-coupled PLL system ($\delta=0.5, g=0.2125, \tau=5.0$). (a) Waveforms of individual oscillators $\sin\theta_1(t)$ (blue) and $\sin\theta_2(t)$ (orange), showing chaotic trajectories remaining nearly synchronized during laminar phases. (b) Long-term evolution of the phase difference $V(t) = (\theta_1-\theta_2)/2$, demonstrating characteristic quiescent laminar states ($V \approx 0$) punctuated by turbulent chaotic bursts.}
\label{fig:timeseries}
\end{figure}

Fig. \ref{fig:timeseries} directly presents the chaotic time-series behavior in the main text. Fig. \ref{fig:timeseries}(a) shows the chaotic trajectory waveforms of individual oscillators, $\sin\theta_1(t)$ and $\sin\theta_2(t)$, which remain almost completely synchronized during laminar intervals. Fig. \ref{fig:timeseries}(b) illustrates the long-term difference $V(t)$, confirming that quiet laminar periods ($V \approx 0$) are randomly interrupted by large-amplitude desynchronization bursts. As detailed in Appendix \ref{sec:appendix_intermittency} and Fig.~\ref{fig:7_app}, long-term statistical evaluations ($T_{\mathrm{max}} = 5 \times 10^5\text{ s}$, $N = 21{,}156$ laminar events) verify that the laminar durations obey $P(T_{\text{lam}}) \propto T_{\text{lam}}^{-3/2}$ with a transverse Lyapunov exponent $\lambda_{\perp} = +7.08 \times 10^{-5}\text{ s}^{-1} \gtrsim 0$.

\subsection{Definition of Latent Space and SVD-Based Effective Dimension}
In this study, the term \textit{latent space} $\mathcal{H}_M \subset \mathbb{R}^M$ refers specifically to the $M$-dimensional state space reconstructed from the continuous scalar phase difference $V(t)$ via a delay-embedding Hankel matrix. This reconstruction maps the functional trajectories of the infinite-dimensional delay differential equations (DDEs) into a finite-dimensional observable subspace governing transverse desynchronization.

For a discrete sampling interval $dt = 0.01$, a sliding temporal window of $W$ data points constructs the $M \times (W-M+1)$ Hankel matrix $\mathbf{H}(t)$:
\begin{equation}
\mathbf{H}(t) = \begin{bmatrix} v_1 & v_2 & \cdots & v_{W-M+1} \\ v_2 & v_3 & \cdots & v_{W-M+2} \\ \vdots & \vdots & \ddots & \vdots \\ v_M & v_{M+1} & \cdots & v_W \end{bmatrix}
\end{equation}
where $v_i = V(t - (W - i)dt)$. Factorizing $\mathbf{H}(t)$ via SVD yields $\mathbf{H}(t) = \mathbf{U} \boldsymbol{\Sigma} \mathbf{V}^T$, where $\mathbf{U}$ and $\mathbf{V}$ are orthogonal matrices and $\boldsymbol{\Sigma} = \text{diag}(\sigma_1, \sigma_2, \dots, \sigma_M)$ contains the singular values sorted in descending order ($\sigma_1 \ge \sigma_2 \ge \dots \ge \sigma_M \ge 0$).

The discrete effective dimension $D_{\text{SVD}}(t)$ counts singular values exceeding a background noise threshold $\epsilon$:
\begin{equation}
D_{\text{SVD}}(t) = \sum_{i=1}^M \mathbb{I}(\sigma_i(t) > \epsilon)
\end{equation}
where $\mathbb{I}(\cdot)$ is the indicator function and $\epsilon = 3\sigma_{\text{noise}}$ is calibrated to the standard deviation of laminar background fluctuations ($V \approx 0$), as discussed in Sec.~\ref{sec:threshold_determination}. Additionally, a continuous effective dimension $D_{\text{cont}}(t)$ based on the inverse participation ratio is evaluated as:
\begin{equation}
D_{\text{cont}}(t) = \frac{\left( \sum_{i=1}^M \sigma_i(t) \right)^2}{\sum_{i=1}^M \sigma_i(t)^2}
\end{equation}
where the inverse participation ratio guarantees that $D_{\text{cont}}(t)$ takes continuous real values strictly bounded within the range $1 \le D_{\text{cont}}(t) \le M$~\cite{mezic2013, susuki2012, Dramac2020}.

\section{Geometric Analysis of Burst Precursors}

\subsection{Temporal Lead-Time and Precursory Detection}
\begin{figure}[htbp]
\centering
\includegraphics[width=0.5\textwidth]{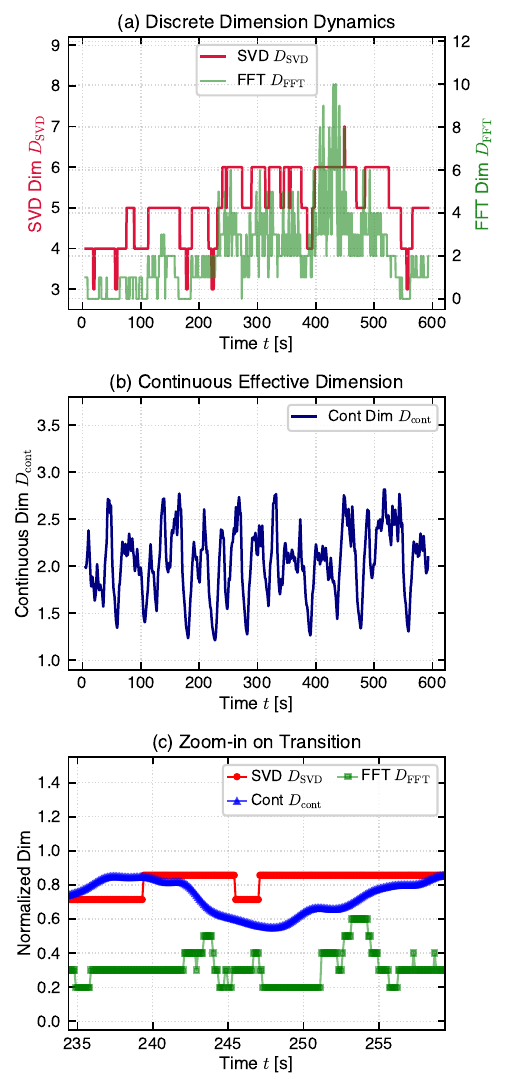} 
\caption{Comparison of temporal evolution between SVD-based effective dimension and FFT-based spectral dimension. (a) Discrete dimensions $D_{\mathrm{SVD}}(t)$ (red) and $D_{\mathrm{FFT}}(t)$ (green). (b) Smooth evolution of continuous effective dimension $D_{\mathrm{cont}}(t)$ ($1 \le D_{\mathrm{cont}} \le M$). (c) Zoom-in view on the transition phase, demonstrating that $D_{\mathrm{SVD}}$ steps up sharply prior to spectral broadening.}
\label{fig:leadtime}
\end{figure}

Fig. \ref{fig:leadtime} compares the temporal evolution of $V(t)$, the SVD-based effective dimension, and the FFT-based spectral dimension across discrete, continuous, and zoomed-in representations.

To establish a benchmark metric, the FFT-based spectral dimension $D_{\text{FFT}}(t)$ is defined as the number of frequency components in the short-time Fourier transform (STFT) amplitude spectrum $\vert{}X(f_k, t)\vert{}$ exceeding threshold $\epsilon_{\text{FFT}}$:
\begin{equation}
D_{\text{FFT}}(t) = \sum_{k=1}^{N_{\text{win}}/2} \mathbb{I}\left( \frac{|X(f_k, t)|}{N_{\text{win}}/2} > \epsilon_{\text{FFT}} \right)
\end{equation}
In Eq. (6), dividing the STFT magnitude $\vert{}X(f_k, t)\vert{}$ by $N_{\text{win}}/2$ normalizes the single-sided discrete Fourier spectrum with respect to the window length $N_{\text{win}}$. Technically, this normalization ensures that $\epsilon_{\text{FFT}}$ acts as an physical amplitude threshold independent of the STFT window size $N_{\text{win}}$, allowing direct physical comparison across different window settings.

As shown in Fig. \ref{fig:leadtime}(a), $D_{\text{FFT}}(t)$ fluctuates rapidly between $0$ and $10$ during chaotic bursts because transient nonlinear harmonics are interpreted as independent spectral dimensions. In contrast, $D_{\text{SVD}}(t)$ remains compact ($4$ to $6$), projecting trajectory snapshots onto the low-dimensional latent manifold and filtering high-frequency noise.

In Fig.~\ref{fig:leadtime}(b), $D_{\text{cont}}(t)$ exhibits smooth, periodic-like fluctuations between $1.3$ and $2.9$ as the singular value spectrum dynamically redistributes energy among spatial-temporal modes. While $D_{\text{cont}}(t)$ captures this continuous energy modulation across $1 \le D_{\text{cont}} \le M$, it does not provide an explicit step-like cutoff for event-driven triggering. In contrast, $D_{\text{SVD}}(t)$ provides a discrete rank-expansion threshold specifically suited for control activation. Crucially, Fig. \ref{fig:leadtime}(c) confirms that $D_{\text{SVD}}(t)$ steps up distinctly prior to any increase in $D_{\text{FFT}}(t)$, establishing structural lead-time in latent space.

\subsection{Mathematical Interpretation via Koopman Operator Framework}
The SVD of the Hankel matrix maps to time-delay embedding, projecting infinite-dimensional DDE dynamics into a finite-dimensional Krylov subspace~\cite{ArbabiMezic2017}. SVD truncates low-variance modes, identifying dominant Koopman observables. Setting trigger threshold $D_{\text{th}}$ translates to a physical rank cutoff where the latent space expands beyond the low-dimensional saddle manifold, detecting incipient transverse instability before spectral broadening~\cite{Ashwin1994}.

\subsection{Estimation of Lead-Time by Cross-Correlation}
To quantify the predictive capability of the SVD-based effective dimension relative to spectral broadening, we compute the normalized cross-correlation function $C(s)$ as a function of time lag $s$:
\begin{equation}
C(s) = \frac{\langle (D_{\text{SVD}}(t) - \bar{D}_{\text{SVD}})(D_{\text{FFT}}(t+s) - \bar{D}_{\text{FFT}}) \rangle}{\sigma_{\text{SVD}} \, \sigma_{\text{FFT}}}
\end{equation}
where $\bar{D}_{\text{SVD}}$ and $\bar{D}_{\text{FFT}}$ denote the temporal averages, $\sigma_{\text{SVD}}$ and $\sigma_{\text{FFT}}$ represent the standard deviations, and $\langle \cdot \rangle$ denotes integration over time $t$. The variable $s$ represents the relative time lag applied to $D_{\text{FFT}}$.

\begin{figure}[htbp]
\centering
\includegraphics[width=0.5\textwidth]{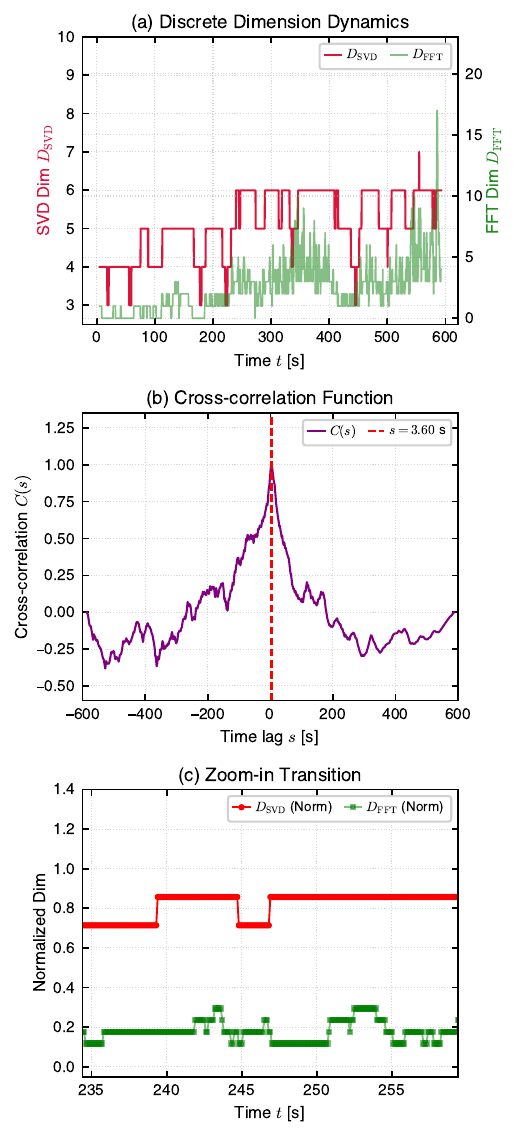} 
\caption{Cross-correlation analysis of dimensional indicators. (a) Discrete dimension dynamics $D_{\mathrm{SVD}}(t)$ and $D_{\mathrm{FFT}}(t)$. (b) Normalized cross-correlation function $C(s)$ versus time lag $s$, displaying a peak at positive lag $s = 3.60\text{ s}$. (c) Zoom-in view illustrating advance dimensional stepping.}
\label{fig:correlation}
\end{figure}

As shown in Fig. \ref{fig:correlation}(b), $C(s)$ exhibits a distinct peak at a positive time lag of $s = 3.60\text{ s}$. While the precise numerical value of $3.60\text{ s}$ quantitatively depends on the choice of threshold parameters ($\epsilon$ and $\epsilon_{\text{FFT}}$), the essential physical conclusion is that the effective dimension in the Hankel-embedded latent space consistently expands prior to observable desynchronization in the time-series waveform. In terms of physical timescales, this positive lag confirms that geometric rank expansion in latent space provides a distinct operational lead-time spanning multiple fundamental oscillation cycles $T_0 = 1/f_0$.

To evaluate the statistical robustness of this precursory behavior across diverse chaotic trajectories, a Monte Carlo simulation ($n=200$ independent uncontrolled trials with random initial phase conditions near the saddle manifold) was conducted. The statistical analysis yields the following characteristics:
\begin{itemize}
    \item \textbf{Mean Lead-Time:} The SVD-based effective dimension preceded the FFT-based spectral dimension by an average of $3.606\text{ s}$ (median $3.100\text{ s}$).
    \item \textbf{Detection Success Rate:} Successful precursory detection (positive lag $s > 0$ prior to physical burst manifestation) occurred in $90.0\%$ of all realizations.
    \item \textbf{Statistical Lead-Time Dispersion:} The standard deviation of $5.409\text{ s}$ reflects the intrinsic trajectory dispersion along unstable transverse directions near the saddle synchronization manifold across independent chaotic realizations. The strong positive skewness guarantees that dimensional growth in latent space acts as a reliable precursor prior to macroscopic physical desynchronization.
\end{itemize}

\section{Proposed Event-Driven Control}

\subsection{Event-Driven Adaptive Delayed Feedback}
The existence of the geometric lead-time identified in Section III enables a forced intervention strategy, where control is applied precisely during the early phase of a burst prior to physical desynchronization. Traditional continuous control remains active even during stable synchronized periods, leading to unintended perturbations of the synchronized manifold and unnecessary energy consumption.

Here, Pyragas-type delayed feedback is specifically chosen over standard proportional state feedback because the Pyragas control term naturally vanishes ($u(t) \to 0$) once phase synchronization ($V(t) = V(t-\tau)$) is restored. It should be emphasized that rather than stabilizing internal unstable periodic orbits of individual oscillators to render them periodic, the present control scheme applies Pyragas-type delayed feedback directly to the transverse coordinate $V(t)$ to stabilize the transverse synchronization manifold $V=0$. Furthermore, compared to dynamically toggling the diffusive coupling parameter $g$---which perturbs the global coupling topology and alters the invariant manifold itself---applying event-driven Pyragas feedback directly to $V(t)$ selectively suppresses incipient transverse growth while preserving the unperturbed autonomous dynamics.

When introduced into the original coupled PLL equations (Eq.~(1)), the control input is applied anti-symmetrically as $+u(t)$ to $\dot{\theta}_1(t)$ and $-u(t)$ to $\dot{\theta}_2(t)$. This directly yields the controlled dynamics for the phase difference $V(t) = (\theta_1 - \theta_2)/2$:
\begin{equation}
\dot{V}(t) = -g \sin(2V(t)) + \frac{1}{2} [ \sin(U_{\tau} - V_{\tau}) - \sin(U_{\tau} + V_{\tau}) ] + u(t)
\end{equation}
where $u(t)$ is the event-driven adaptive delayed feedback defined as:
\begin{equation}
u(t) = \begin{cases} -K [V(t) - V(t-\tau)] & \text{if } dim_{\text{eff}}(t) \ge D_{\text{th}} \\ 0 & \text{if } dim_{\text{eff}}(t) < D_{\text{th}} \end{cases}
\end{equation}
By directly compensating for deviations from the saddle synchronization manifold $V=0$, the controller suppresses the transverse instability detected by the SVD-based effective dimension.

\subsection{Systematic Determination of Cutoff Thresholds}
\label{sec:threshold_determination}
In this study, the trigger threshold $D_{\text{th}} = 1.794$ is determined as a physical rank cutoff that identifies the expansion of the active observable subspace within the Koopman operator framework~\cite{ArbabiMezic2017}. While the trajectory remains confined near the laminar saddle synchronization manifold, the Hankel matrix maintains a low-rank structure dominated by $D_{\text{cont}} \approx 1.3$. When transverse instability is excited, higher-order singular values cross the noise floor, expanding the required subspace dimension. System parameters are summarized in Table~\ref{tab:params}.

\begin{table}[htbp]
\renewcommand{\arraystretch}{1.3}
\caption{Simulation and Analysis Parameters}
\label{tab:params}
\centering
\begin{tabular}{l||l|l}
\hline
\bfseries Symbol & \bfseries Parameter Name & \bfseries Value \\
\hline\hline
$\delta$ & Natural frequency deviation & 0.5 \\
$g$ & Coupling strength & 0.2125 \\
$\tau$ & Normalized time delay & 5.0 \\
$dt$ & Oversampling step & 0.01 \\
$D_{\text{th}}$ & Effective dimension threshold & 1.794 \\
$K$ & Control gain & 1.2 \\
\hline
\end{tabular}
\end{table}

The systematic determination of the thresholds ($\epsilon$ and $D_{\text{th}}$) follows a quantitative three-step procedure:
\begin{enumerate}
    \item \textbf{Noise Floor Characterization ($3\sigma$ Rule):} The SVD noise-floor threshold $\epsilon$ is quantitatively evaluated during the laminar regime ($\vert{}V\vert{} \approx 0$). Linear background fluctuations exhibit a standard deviation $\sigma_{\text{noise}}$, establishing the truncation cutoff $\epsilon = 3\sigma_{\text{noise}}$ to filter out non-essential residual modes.
    \item \textbf{Statistical Boundary for $D_{\text{th}}$:} To monitor the PDF of $\text{dim}_{\text{eff}}(t)$ during stable laminar operation, $D_{\text{th}}$ is set strictly above baseline noise fluctuations.
    \item \textbf{Subspace Rank Expansion Detection:} When $\text{dim}_{\text{eff}}(t)$ exceeds $D_{\text{th}} = 1.794$, it detects a structural rank expansion in the Hankel embedding space, signifying that unstable transverse modes have been activated and require a higher-dimensional observable subspace to capture the trajectory prior to physical desynchronization.
\end{enumerate}

\begin{figure}[htbp]
\centering
\includegraphics[width=0.5\textwidth]{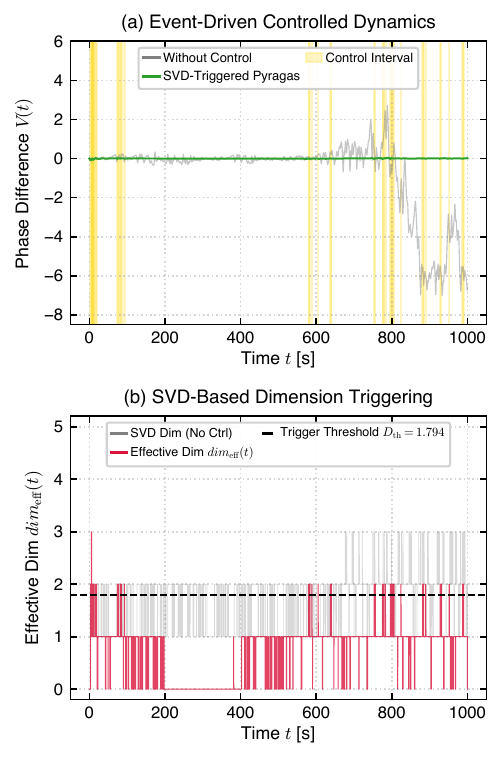} 
\caption{Performance of SVD-triggered event-driven control. (a) Phase difference $V(t)$ under uncontrolled (gray) and controlled (green) states. (b) Dynamic dimension $\mathrm{dim}_{\mathrm{eff}}(t)$, triggering control when exceeding $D_{\mathrm{th}} = 1.794$.}
\label{fig:control}
\end{figure}

\section{Performance Evaluation}

\subsection{Suppression Efficacy and Predictive Advantage}
As demonstrated in Fig. \ref{fig:control}, the proposed SVD-triggered control scheme effectively suppresses intermittent chaotic bursts immediately after detection. By directly tracking rank expansion in the Hankel embedding space, the SVD-based effective dimension provides a computationally efficient, data-driven indicator for event-driven control. Proactive stabilization driven by geometric precursors in latent space is physically superior to conventional reactive control, which responds only after macroscopic oscillations have already manifested in scalar state variables. Our results demonstrate that tracking the effective dimension of latent dynamics provides a reliable indicator for detecting and arresting impending transverse instability.

\begin{figure}[htbp]
\centering
\includegraphics[width=0.5\textwidth]{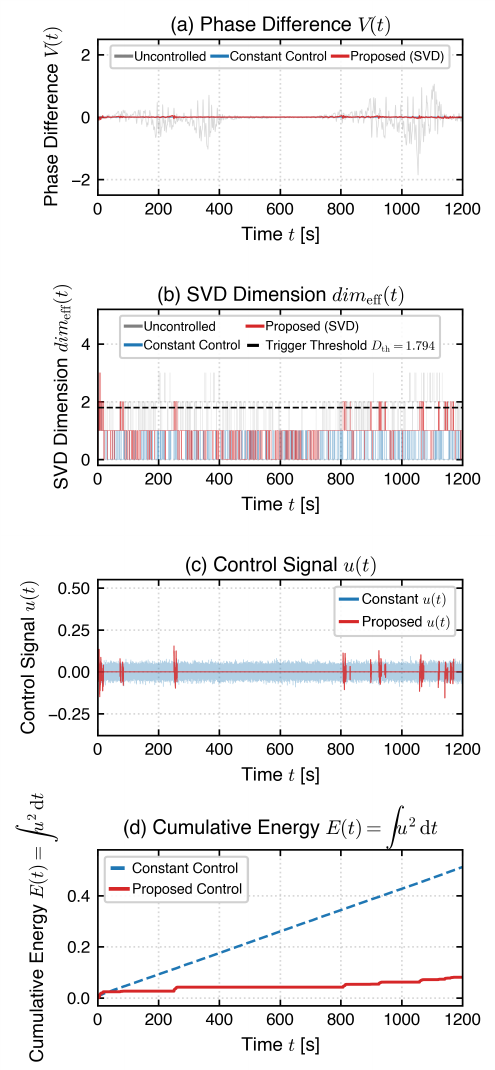} 
\caption{Control effort evaluation via cumulative energy $E(t) = \int u^2 dt$. (a) Phase difference $V(t)$. (b) Dynamic dimension $dim_{\text{eff}}(t)$. (c) Control signals $u(t)$. (d) Cumulative energy $E(t)$: continuous Pyragas control (blue) increases linearly, whereas SVD-triggered control (red) maintains zero consumption during laminar states.}
\label{fig:energy}
\end{figure}

\subsection{Energy Efficiency and Stabilization}
Fig. \ref{fig:energy} quantifies energy savings. Continuous Pyragas control counteracts noise continuously, increasing $E(t)$ linearly. The SVD-triggered scheme remains inactive ($u=0$) during laminar periods, consuming energy only in discrete steps during necessary interventions.

\section{Conclusion}
This paper has presented a data-driven framework for predicting and suppressing on-off intermittent bursts in delay-coupled PLLs by tracking dynamic transitions of the effective dimension in latent space (defined as the finite-dimensional state space reconstructed via Hankel matrix embedding).

While works by Mezi{\'c} and Susuki~\cite{mezic2013, susuki2012} established the power of spectral Koopman operator theory for analyzing global fluid and power system dynamics, this study demonstrates that monitoring the effective dimension of the transverse latent subspace allows for sensitive detection of early trajectory escape away from the saddle synchronization manifold. In the studied coupled PLL system, SVD-based detection captures the dynamic dimensional transition from a low-dimensional state ($D_{\text{cont}} \approx 1.3$) representing transverse laminar confinement to a higher-dimensional state ($dim_{\text{eff}} \ge 3$) associated with chaotic bursting.

The primary contribution of this work is the realization of a proactive, event-driven delayed feedback control strategy. By quantifying the lead-time through cross-correlation analysis, we established that dimensional growth in latent space consistently precedes observable spectral broadening. Triggering Pyragas-type feedback upon detecting rank expansion allows the controller to act during the incipient stage of transverse instability, substantially reducing cumulative control energy while preserving the unperturbed autonomous dynamics near the saddle manifold.

The methodology of identifying dynamic transitions in effective dimension offers a practical data-driven strategy for detecting precursor instabilities in delay-coupled systems.

\section*{Acknowledgments}
This work was partially supported by JSPS KAKENHI Grant Number JP26K07513. The authors express sincere gratitude to Professor Igor Mezi{\'c} (UCSB) and Professor Yoshihiko Susuki (Kyoto University) for insightful discussions regarding Koopman Operator Theory and Hankel-DMD alignment.

\appendix

\section{Quantitative Statistical Verification of On-Off Intermittency}
\label{sec:appendix_intermittency}

\begin{figure}[htbp]
\centering
\includegraphics[width=0.5\textwidth]{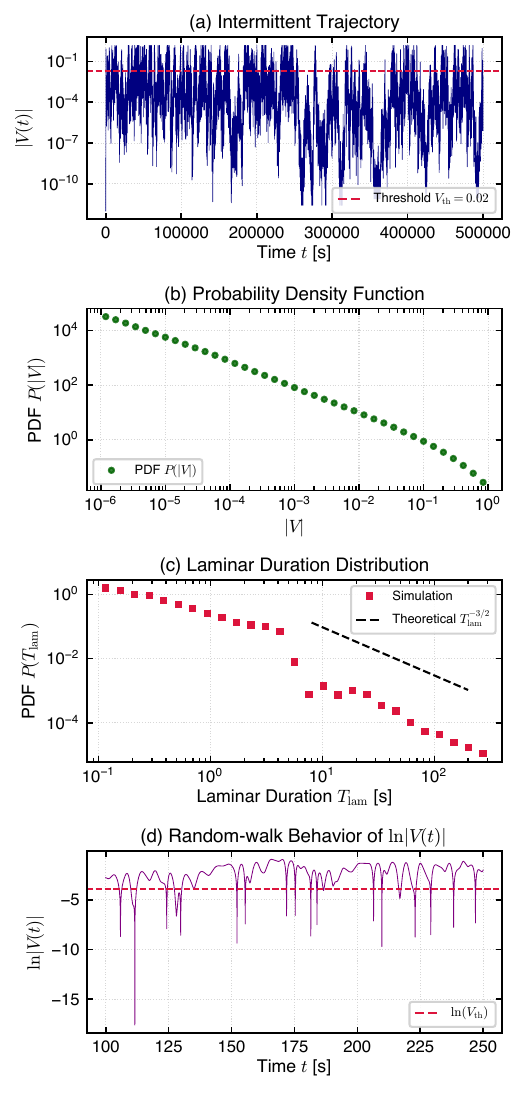}
\caption{Quantitative statistical evidence of on-off intermittency in the mutually coupled PLL system ($g = 0.2125, \delta = 0.5, \tau = 5.0, T_{\mathrm{max}} = 5 \times 10^5\text{ s}$). 
(a) Logarithmic time series of $\vert{}V(t)\vert{}$ highlighting laminar trapping ($10^{-5}$--$10^{-3}$) and burst events exceeding $V_{\mathrm{th}} = 0.02$. 
(b) Probability density function $P(\vert{}V\vert{})$ showing power-law scaling $P(\vert{}V\vert{}) \propto \vert{}V\vert{}^{-1+\eta}$ near $\vert{}V\vert{} \to 0$. 
(c) Distribution of laminar durations $P(T_{\mathrm{lam}})$ for $N = 21{,}156$ events (red squares). Beyond the delay memory timescale ($T_{\mathrm{lam}} \ge 8.0\text{ s}$), the fitted asymptotic scaling exponent is $\alpha = -1.538 \pm 0.122$ ($R^2 = 0.9468$), exhibiting remarkable agreement with the theoretical asymptotic $-3/2$ scaling reference line (black dashed line). 
(d) Diffusive random-walk trajectories of $\ln\vert{}V(t)\vert{}$ below $\ln(V_{\mathrm{th}})$, governed by positive drift $\lambda_{\perp} = +7.08 \times 10^{-5}\text{ s}^{-1}$ and diffusion $D = 1.68\text{ s}^{-1}$.}
\label{fig:7_app}
\end{figure}

To complement Section II.A of the main manuscript, this appendix provides explicit, quantitative numerical evidence confirming that the desynchronization bursts observed in the mutually coupled phase-locked loops (PLLs) strictly correspond to deterministic on-off intermittency.

All statistical quantities presented below were evaluated from long-term numerical integrations with a total integration time of $T_{\mathrm{max}} = 5 \times 10^5\text{ s}$ ($5 \times 10^7$ integration steps with $dt = 0.01$) near the critical desynchronization threshold ($g = 0.2125, \delta = 0.5, \tau = 5.0$). Fig.~\ref{fig:7_app} summarizes the four foundational quantitative statistical indicators:

\begin{enumerate}
    \item \textbf{Intermittent Time Series (Fig.~\ref{fig:7_app}a):} On a logarithmic scale, the phase difference $\vert{}V(t)\vert{}$ remains trapped in the laminar state near the numerical noise floor ($10^{-5}$--$10^{-3}$) for long intervals before abruptly expanding into large-scale chaotic bursts exceeding the threshold $V_{\mathrm{th}} = 0.02$. Continuous phase-wrapping via $\mathrm{arctan2}$ ensures rigorous tracking across phase slips over the full integration interval $T_{\mathrm{max}} = 5 \times 10^5\text{ s}$.
    
    \item \textbf{Probability Density Function (Fig.~\ref{fig:7_app}b):} The probability density function $P(\vert{}V\vert{})$ exhibits a singular power-law distribution near the synchronization manifold ($\vert{}V\vert{} \to 0$) alongside a heavy tail in the large-amplitude burst regime, described by $P(\vert{}V\vert{}) \propto \vert{}V\vert{}^{-1+\eta}$ with an exponent parameter $\eta = 2\lambda_{\perp}/D \approx 8.41 \times 10^{-5}$.
    
    \item \textbf{Laminar Phase Duration Distribution (Fig.~\ref{fig:7_app}c):} From a large statistical sample of $N = 21{,}156$ laminar events defined by $\vert{}V(t)\vert{} < V_{\mathrm{th}} = 0.02$, the distribution of laminar durations $T_{\mathrm{lam}}$ was computed. For short durations ($T_{\mathrm{lam}} < 8.0\text{ s}$), the distribution is governed by deterministic transient memory due to the delay feedback ($\tau = 5.0\text{ s}$). In the true asymptotic diffusion regime ($T_{\mathrm{lam}} \ge 8.0\text{ s}$), a linear least-squares fit yields a scaling exponent of $\alpha = -1.538 \pm 0.122$ ($R^2 = 0.9468$), which is in excellent agreement with the theoretical asymptotic scaling exponent of $-3/2$ ($T_{\mathrm{lam}}^{-1.500}$). Robustness across $V_{\mathrm{th}} \in [0.01, 0.04]$ was also quantitatively verified.
    
    \item \textbf{Diffusive Random-Walk Dynamics (Fig.~\ref{fig:7_app}d):} In the laminar regime ($\vert{}V(t)\vert{} < V_{\mathrm{th}}$), the log-amplitude $y(t) = \ln\vert{}V(t)\vert{}$ behaves as a biased random walk described by $dy(t) = \lambda_{\perp} dt + \sqrt{D} dW(t)$. Quantitative evaluation from long-term laminar trajectories yields an average transverse drift rate (transverse Lyapunov exponent, evaluated as $\lambda_{\perp} = \lim_{t \to \infty} \frac{1}{t} \ln \frac{\vert{}V(t)\vert{}}{\vert{}V(0)\vert{}}$) extremely close to zero, $\lambda_{\perp} = +7.08 \times 10^{-5}\text{ s}^{-1}$, accompanied by a diffusion coefficient of $D = 1.68\text{ s}^{-1}$. The condition $\lambda_{\perp} \gtrsim 0$ quantitatively confirms operation in the immediate supercritical vicinity of the blowout bifurcation threshold, where diffusive fluctuations regularly drive the latent dynamics into macroscopic chaotic desynchronization bursts.
\end{enumerate}

\end{document}